\documentclass[prl,final,twocolumn,showpacs,floatfix]{revtex4}%
\usepackage{amsfonts}
\usepackage{amsmath}
\usepackage{amssymb}
\usepackage{graphicx}%
\begin{document}
\title{Lattice effective field theory calculations for $A=3,4,6,12$ nuclei}
\author{{Evgeny Epelbaum$^{a,b}$, Hermann~Krebs$^{b,a}$, Dean~Lee$^{c,b}$,
Ulf-G.~Mei{\ss }ner$^{b,a,d}$} \linebreak}
\affiliation{$^{a}$Institut f\"{u}r Kernphysik (IKP-3) and J\"{u}lich Center for Hadron
Physics, \linebreak Forschungszentrum J\"{u}lich, D-52425 J\"{u}lich, Germany
\linebreak$^{b}$Helmholtz-Institut f\"{u}r Strahlen- und Kernphysik (Theorie)
and \linebreak Bethe Center for Theoretical Physics, Universit\"{a}t Bonn,
D-53115 Bonn, Germany \linebreak$^{c}$Department of Physics, North Carolina
State University, \linebreak Raleigh, NC 27695, USA \linebreak$^{d}$Institute
for Advanced Simulation (IAS), Forschungszentrum J\"{u}lich, D-52425
J\"{u}lich, Germany}

\begin{abstract}
We present lattice results for the ground state energies of tritium, helium-3,
helium-4, lithium-6, and carbon-12 nuclei. \ Our analysis includes
isospin-breaking, Coulomb effects, and interactions up to
next-to-next-to-leading order in chiral effective field theory.

\end{abstract}
\maketitle
\preprint{ }

Several ab initio approaches have been used to calculate the properties of
various few- and many-nucleon systems. \ Some recent work includes the no-core
shell model
\cite{Forssen:2004dk,Nogga:2005hp,Stetcu:2006ey,Navratil:2007we,Maris:2008ax},
constrained-path
\cite{Pieper:2002ne,Pieper:2004qh,Marcucci:2008mg,Pieper:2009aa} and
fixed-node \cite{Chang:2004sj,Gezerlis:2007fs} Green's function Monte Carlo,
auxiliary-field diffusion Monte Carlo
\cite{Gandolfi:2007hs,Gandolfi:2009fj,Gandolfi:2009nq}, and coupled cluster
methods \cite{Wloch:2005za,Hagen:2007hi,Hagen:2008iw}. \ The diversity of
methods is useful since each technique has its own computational scaling,
systematic errors, and range of accessible problems. \ Furthermore, quantities
not directly measured in experiments can be benchmarked with calculations
using other methods.

Another ab initio approach in the recent literature is lattice effective field
theory. \ This method combines the theoretical framework of effective field
theory (EFT) with numerical lattice methods. \ When compared with other
methods it is unusual in that all systematic errors are introduced up front
when defining the truncated low-energy effective theory. \ This eliminates
approximation errors tied with a specific calculational tool, physical system,
or observable. \ By including higher-order interactions in the low-energy
effective theory, one can reasonably expect systematic and systemic
improvement for all low-energy observables. \ The approach has been used to
simulate nuclear matter \cite{Muller:1999cp} and neutron matter
\cite{Lee:2004qd,Lee:2004si,Abe:2007fe,Borasoy:2007vk,Epelbaum:2008vj,Wlazlowski:2009yi}%
. \ The method has also been applied to nuclei with $A\leq4$ in pionless EFT
\cite{Borasoy:2005yc} and chiral EFT \cite{Borasoy:2006qn,Epelbaum:2009zs}.
\ A review of lattice effective field theory calculations can be found in
Ref.~\cite{Lee:2008fa}.

In this letter we present the first lattice calculations for lithium-6 and
carbon-12 using chiral effective field theory. \ We address a fundamental
question in the nuclear theory community: \ Can effective field theory be
applied to nuclei beyond the very lightest? \ While there are several
calculations that probe this question using interactions derived from chiral
effective field theory, we present the first calculations posed and computed
entirely within the framework of effective field theory. \ Our results show
that lattice-regularized effective field theory can be applied to the ground
state of carbon-12. \ Furthermore there is a clear path towards larger nuclei
and nuclear matter. \ We also describe the first lattice calculations to
include isospin-breaking and Coulomb interactions, and compute the energy
splitting between helium-3 and the triton. \ Our discussion focuses on new
features of the calculation and new results. \ A complete description of the
calculational method is contained in a separate paper \cite{Epelbaum:2010a}.

The low-energy expansion in effective field theory is organized in powers of
$Q$/$\Lambda$, where $Q$ is the low momentum scale associated with external
nucleon momenta or the pion mass, and $\Lambda$ is the high momentum scale at
which the effective theory breaks down. \ Some reviews of chiral effective
field theory can be found in
Ref.~\cite{vanKolck:1999mw,Bedaque:2002mn,Epelbaum:2005pn,Epelbaum:2008ga}.
\ At leading order (LO) in the Weinberg power counting scheme the
nucleon-nucleon effective potential contains two independent contact
interactions and instantaneous one-pion exchange. \ As in previous lattice
studies we make use of an \textquotedblleft improved\textquotedblright%
\ leading-order\ action. \ This improved leading-order action is treated
completely non-perturbatively, while higher-order interactions are included as
a perturbative expansion in powers of $Q/\Lambda$.

We use the improved LO$_{3}$ lattice action introduced in
Ref.~\cite{Epelbaum:2008vj} with spatial lattice spacing $a=(100$%
~MeV$)^{-1}=1.97$~fm and temporal lattice spacing $a_{t}=(150$~MeV$)^{-1}%
=1.32$~fm. \ The interactions provide a good description of the neutron-proton
$S$-wave and $P$-wave phase shifts at low energies as well as the $S$-$D$
mixing angle. \ Plots of the scattering data for the LO$_{3}$ lattice action
can be found in Ref.~\cite{Epelbaum:2008vj}. \ The corrections at
next-to-leading order (NLO) and next-to-next-to-leading order (NNLO) are
calculated using perturbation theory. \ A description of these interactions on
the lattice is documented in Ref.~\cite{Epelbaum:2009zs}.

At NLO there are corrections to the two leading-order coefficients and seven
additional unknown coefficients for operators with two powers of momentum.
\ These nine coefficients are determined by fitting to the neutron-proton
$S$-wave and $P$-wave phase shifts and $S$-$D$ mixing angle at low energies.
\ At NNLO there are two additional cutoff-dependent coefficients associated
with three-nucleon interactions. \ These are parameterized by two
dimensionless coefficients $c_{D}$ and $c_{E}$, corresponding with the
three-nucleon one-pion exchange diagram and three-nucleon contact interaction
respectively. \ We constrain $c_{E}$ by requiring that the triton energy
equals the physical value of $-8.48$~MeV. \ However the parameter $c_{D}$ is
relatively unconstrained by low-energy phenomena such as the deuteron-neutron
spin-doublet phase shifts. \ Currently we are investigating other methods for
constraining $c_{D}$, including one recent suggestion to determine $c_{D}$
from the triton beta decay rate \cite{Gazit:2008ma}. \ In this analysis we
simply use the estimate $c_{D}\sim O(1)$ and check the dependence of
observables upon changes in $c_{D}$.

In addition to isospin-symmetric interactions, we also include
isospin-breaking (IB) and electromagnetic (EM) interactions. \ Isospin
violation in effective field theory has been addressed extensively in the
literature
\cite{vanKolck:1996rm,vanKolck:1997fu,Epelbaum:1999zn,Walzl:2000cx,Friar:2003yv,Epelbaum:2005fd}%
. \ In the counting scheme proposed in Ref.~\cite{Epelbaum:2005fd}, the
isospin-breaking one-pion exchange interaction and Coulomb potential are
numerically the same size as $O(Q^{2}/\Lambda^{2})$ corrections at NLO. \ On
the lattice we treat the Coulomb potential in position space with the usual
$\alpha_{\text{EM}}/r$ dependence. \ However this definition is singular for
two protons on the same lattice site and requires short-distance
renormalization via a proton-proton contact interaction. \ In this study we
include all possible contact interactions, namely interactions for
neutron-neutron, proton-proton, spin-singlet neutron-proton, and spin-triplet
neutron-proton. \ The two neutron-proton contact interactions are already
included at NLO and determined from neutron-proton scattering. \ The other two
coefficients are determined from fitting to $S$-wave phase shifts for
proton-proton scattering and the neutron-neutron scattering length. \ Details
of this calculation will be presented in a separate paper
\cite{Epelbaum:2010a}.

\begin{figure}
[ptb]
\begin{center}
\includegraphics[
height=1.7772in,
width=2.0366in
]%
{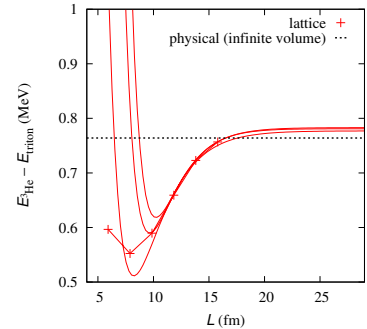}%
\caption{Plot of the energy difference between helium-3 and the triton as a
function of periodic cube length.}%
\label{triton_he3_difference}%
\end{center}
\end{figure}

The first results we present are for helium-3 and the triton. \ The
three-nucleon system is sufficiently small that we can use iterative
sparse-matrix eigenvector methods to compute helium-3 and the triton on cubic
periodic lattices. \ We consider cubes with side lengths $L$ up to $16$~fm and
extract the infinite volume limit using the asymptotic parameterization
\cite{Luscher:1985dn}, $E(L)\approx E(\infty)-ce^{-L/L_{0}}/L$. \ While the
triton energy at infinite volume is used to set the unknown coefficient
$c_{E}$, the energy splitting between helium-3 and the triton is a prediction
that can be compared with experiment. \ The energy difference between helium-3
and the triton is plotted in Fig. \ref{triton_he3_difference} as a function of
cube length. \ We show several different asymptotic fits using different
subsets of data points. \ To the order at which we are working there is no
dependence of the energy splitting upon the value of $c_{D}$. \ Our
calculations at next-to-next-to-leading order give a value of $0.780$~MeV with
an infinite-volume extrapolation error of $\pm0.003$~MeV. \ To estimate other
errors we take into account an uncertainty of $\pm1$~fm in the neutron
scattering length and a $5\%$ relative uncertainty in our lattice fit of the
splitting between neutron-proton and proton-proton phase shifts at low
energies. \ Our final result for the energy splitting with error bars is
$0.78(5)$~MeV. \ This agrees well with the experimental value of $0.76$~MeV.%

For systems with more than three nucleons, we use projection Monte Carlo with
auxiliary fields and extract the properties of the ground state using
Euclidean-time projection. \ The transfer matrix, $M$, is the normal-ordered
exponential of the Hamiltonian over one temporal lattice spacing. \ As in
previous lattice Monte Carlo simulations we first define a transfer matrix
$M_{\text{SU(4)}\not \pi }$ which is invariant under Wigner's SU(4) symmetry
rotating all spin and isospin components of nucleons. \ This transfer matrix
acts as an approximate low-energy filter that happens to be computationally
inexpensive. \ Starting from a Slater determinant of free-particle standing
waves, $\left\vert \Psi^{\text{free}}\right\rangle $, we construct the trial
state $\left\vert \Psi(t^{\prime})\right\rangle $ by successive
multiplication,%
\begin{equation}
\left\vert \Psi(t^{\prime})\right\rangle =\left(  M_{\text{SU(4)}\not \pi
}\right)  ^{L_{t_{o}}}\left\vert \Psi^{\text{free}}\right\rangle ,
\label{L_t_o}%
\end{equation}
where $t^{\prime}=L_{t_{o}}a_{t}$ and $L_{t_{o}}$ is the number of
\textquotedblleft outer\textquotedblright\ time steps. \ The trial function
$\left\vert \Psi(t^{\prime})\right\rangle $ is then used as the starting point
for the calculation. \ The amplitude $Z(t)$ is defined as%
\begin{equation}
Z(t)=\left\langle \Psi(t^{\prime})\right\vert \left(  M_{\text{LO}}\right)
^{L_{t_{i}}}\left\vert \Psi(t^{\prime})\right\rangle , \label{L_t_i}%
\end{equation}
where $t=L_{t_{i}}a_{t}$ and $L_{t_{i}}$ is the number of \textquotedblleft
inner\textquotedblright\ time steps. \ The transient energy $E(t)$ is
proportional to the logarithmic derivative of $Z(t)$, and the ground state
energy is given by the limit of $E(t)$ as $t\rightarrow\infty$. \ Each of the
transfer matrices are functions of the auxiliary fields and pion fields, and
the Monte Carlo integration over field configurations is performed using
hybrid Monte Carlo. \ Contributions due to NLO\ and NNLO\ interactions,
isospin breaking (IB), and electromagnetic interactions (EM) are incorporated
using perturbation theory.

\begin{figure}
[ptb]
\begin{center}
\includegraphics[
height=1.9242in,
width=3.1427in
]%
{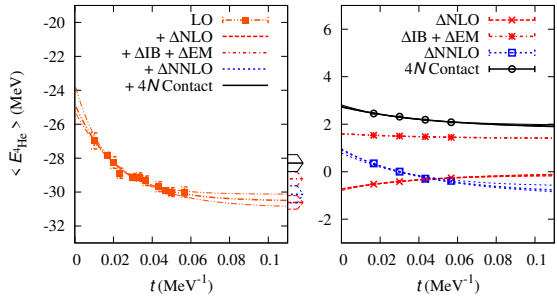}%
\caption{Ground state energy for helium-4 as a function of Euclidean time
projection.}%
\label{alpha}%
\end{center}
\end{figure}

In Fig.~\ref{alpha} we show lattice results for the ground state of helium-4
in a periodic cube of length $9.9$~fm. \ For the numerical extrapolation in
Euclidean time we use the decaying exponential functions described in
Ref.~\cite{Epelbaum:2009zs}. \ The plot on the left shows the contributions
from leading-order and higher-order contributions added cumulatively. \ The
plot on the right shows the higher-order corrections separately. \ For each
case we show the best fit as well as the one standard-deviation bound. \ We
estimate this bound by generating an ensemble of fits determined with added
random Gaussian noise proportional to the error bars of each data point and
varying the number of fitted data points. \ These results are similar to those
found in Ref.~\cite{Epelbaum:2009zs} using the LO$_{2}$ action. \ For
$c_{D}=1$ we get $-30.5(4)$~MeV at LO, $-30.6(4)$~MeV at NLO, $-29.2(4)$~MeV
at NLO with IB and EM corrections, and $-30.1(5)$~MeV at NNLO. \ The helium-4
energy decreases $0.4(1)$~MeV for each unit increase in $c_{D}$.%

The size of the corrections at NNLO gives an estimate of the remaining error
from higher-order terms in the effective field theory expansion. \ Given our
cutoff momentum scale of $\Lambda=\pi/a=314$~MeV, an error of $1$ to $2$~MeV
is consistent with the expected size of higher-order contributions.
\ Interactions at higher order than NNLO are beyond the scope of the current
calculation. However if it happens that the higher-order effects are most
important when all four nucleons are in close proximity, then we should see
universal behavior which can be reproduced by an effective four-nucleon
contact interaction. \ We test this universality hypothesis by introducing an
effective four-nucleon contact interaction tuned to reproduce the physical
helium-4 energy of $-28.3$~MeV. \ The contribution of this interaction in
helium-4 is shown in Fig.~\ref{alpha}.

\begin{figure}
[ptb]
\begin{center}
\includegraphics[
height=1.9233in,
width=3.1427in
]%
{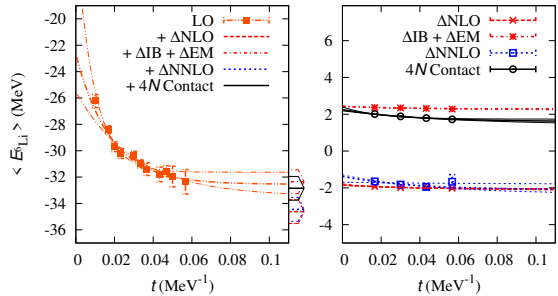}%
\caption{Ground state energy for lithium-6 as a function of Euclidean time
projection.}%
\label{lithium6}%
\end{center}
\end{figure}

In Fig.~\ref{lithium6} we show lattice results for the ground state of
lithium-6 in a periodic cube of length $9.9$~fm. \ For $c_{D}=1$ we get
$-32.6(9)$~MeV at LO, $-34.6(9)$~MeV at NLO, $-32.4(9)$~MeV at NLO with IB and
EM corrections, and $-34.5(9)$~MeV at NNLO. $\ $Adding the contribution of the
effective four-nucleon interaction to the NNLO result gives $-32.9(9)$~MeV.
\ This lies within error bars of the physical value $-32.0$~MeV. \ However we
expect some overbinding due to the finite periodic volume, and the deviation
of $0.9$~MeV is consistent with the expected size of the finite volume
correction. \ Further calculations at varying volumes will be needed to
determine this volume dependence. \ Without the effective four-nucleon
interaction, the lithium-6 energy decreases $0.7(1)$~MeV for each unit
increase in $c_{D}$. \ With the effective four-nucleon interaction the
lithium-6 energy decreases $0.35(5)$~MeV per unit increase in $c_{D}$.%

\begin{figure}
[ptb]
\begin{center}
\includegraphics[
height=1.9233in,
width=3.1427in
]%
{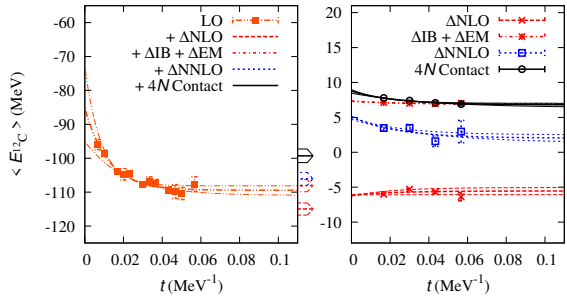}%
\caption{Ground state energy for carbon-12 as a function of Euclidean time
projection.}%
\label{carbon12}%
\end{center}
\end{figure}

In Fig.~\ref{carbon12} we show lattice results for the ground state of
carbon-12 in a periodic cube of length $13.8$~fm. \ For $c_{D}=1$ we get
$-109(2)$~MeV at LO, $-115(2)$~MeV at NLO, $-108(2)$~MeV at NLO with IB and EM
corrections, and $-106(2)$~MeV at NNLO. $\ $Adding the contribution of the
effective four-nucleon interaction to the NNLO result gives $-99(2)$~MeV.
\ This is an overbinding of $7\%$ from the physical value, $-92.2$~MeV. \ We
note that an overbinding of $7\%$ is actually a reasonable estimate of the
finite volume correction for carbon-12 in a periodic box of length $13.8$~fm.
This suggests that at infinite volume the error is significantly smaller than
$7\%$. \ Further calculations at varying volumes will be needed to measure the
volume dependence. \ Without the effective four-nucleon interaction, the
carbon-12 energy decreases $1.7(3)$~MeV for each unit increase in $c_{D}$.
\ With the effective four-nucleon interaction the carbon-12 energy decreases
$0.3(1)$~MeV per unit increase in $c_{D}$.%

The results for lithium-6 and carbon-12 appear to confirm the universality
hypothesis regarding higher-order interactions. The much reduced dependence
upon on $c_{D}$ is also consistent with the universality hypothesis. \ The
effective four-nucleon contact interaction can be viewed as absorbing the
dependence on $c_{D}$. \ We note a recent related paper on the renormalization
group evolution of higher-nucleon interactions \cite{Jurgenson:2009qs}. The
accuracy of these lattice calculations are competitive with recent
calculations obtained using other ab initio methods. \ Constrained-path
Green's function Monte Carlo calculations and no-core shell model calculations
have an accuracy of $1\%-2\%$ in energy for nuclei $A\leq12$. \ Coupled
cluster calculations without three-nucleon interactions are accurate to within
$1$~MeV per nucleon for medium mass nuclei.\ \ Future lattice studies should
look at probing large volumes, including higher-order effects, and decreasing
the lattice spacing.

Lattice effective field theory combines the generality of effective field
theory with the flexibility of lattice methods. \ The computational scaling of
the calculations presented here indicates that larger systems with more
nucleons should be possible. \ By applying different lattice boundary
conditions in the spatial and temporal directions, it is possible to probe
nuclear systems of many different varieties: \ few-body and many-body systems;
zero temperature and nonzero temperature; nuclear matter, neutron matter, and
asymmetric nuclear matter.

\textit{Partial financial support provided by the Deutsche
Forschungsgemeinschaft, Helmholtz Association, U.S. Department of Energy, and
EU HadronPhysics2 Project. \ Computational resources provided by the
J\"{u}lich Supercomputing Centre.}

\end{document}